\documentstyle[12pt, psfig]{article}
\newcommand{\beq}{\begin{equation}}
\newcommand{\eeq}{\end{equation}}
\newcommand{\pd}{\partial}
\renewcommand{\k}{\kappa}
\begin{document}
\begin{titlepage}
\hfill BI-TP-96/50

\hfill November 1996

\begin{center}
{\large \bf TOPOLOGY IN 4D SIMPLICIAL QUANTUM GRAVITY} 

\vspace{1.7cm}

{\sl S. Bilke\footnotemark[1], Z. Burda\footnotemark[2] and B. Petersson\footnotemark[1]}

\vspace{0.6cm}

\footnotemark[1]
Fakult\"{a}t f\"{u}r Physik
Universit\"{a}t Bielefeld, \\
Postfach 10 01 31, Bielefeld 33501, Germany\\

\vspace{0.3cm}

\footnotemark[2]
Institute of Physics Jagellonian University,\\
ul. Reymonta 4, PL-30 059, Krak\'{o}w~16, Poland\\

\vfill
{\bf Abstract}
\end{center}

\noindent
We simulate $4d$ simplicial gravity for three topologies 
$S^4$, $S^3 \times S^1$ and $S^1 \times S^1 \times S^1 \times S^1$
and show that the free energy for these three fixed topology 
ensembles is the same in the thermodynamic limit 
$N_4 \rightarrow \infty$. We show, that the next-to-leading order 
corrections, at least away from the critical point, can be described
by kinematic sources.
\vfill
\end{titlepage}

\section{Introduction}
It is not a priori clear, whether in a path-integral formulation of
quantum gravity the sum over metrics should also run over topologies.
In a theory containing topology fluctuations only those topologies
contribute to the sum in the thermodynamic limit, which maximize
the extensive part of the free energy. Other contributions are
exponentially suppressed. If one believes that {\em all} topology
excitations should be present in the continuum theory, the bulk
volume contribution to the free energy should be independent of 
topology. An explicit example of such a theory is provided by 
two dimensional quantum gravity.
Here the Einstein-Hilbert action is of purely topological nature.
It can  be shown explicitly that the coefficent $\mu _c$ 
of the leading (extensive) part of the entropy  does  not depend 
on the topology. The coefficient of the logarithmic correction 
is $\gamma -3$, where $\gamma $ is the so-called surface susceptibility
exponent. The exponent $\gamma $, does, however, depend linearly
on the genus of the surface. The linear dependence leads to the
double scaling limit at which one can reduce the number of 
non-perturbative modes of the theory with fluctuating topology to the 
solutions of the Painlev{\'e} II equation.
In four dimensions the situation is more complicated. At present, 
no classification of topologies is known. Therefore the
topological part in the action is unknown. Simplicial quantum gravity
allows us to sum over geometries with fixed topology.  We investigate
numerically three different topologies and show that  in these
cases, up to the leading order, the free energy does not depend
on the topology. We observe a topology dependence in the next to
leading order.  We study those volume corrections and analyze their 
sources at the kinematic bounds. Some of our results have already been
presented at Lattice 96 \cite{bibkp96}. 

\section{Definitions}
The partition function of simplicial quantum gravity in the grand canonical 
ensemble is
\beq
Z(\k _2, \k _4) = \sum _{\cal T} \frac{1}{C(T)} e^{- \k _4 N_4 + \k _2 N_2},
\label{gcan}
\eeq
where the first summation is over all $4d$-simplicial manifolds $\cal T$ 
with fixed topology \cite{aj1}. 
The parameter $\k_4$ is proportional to the cosmological 
constant and $\k_2$ is a linear combination of the inverse of 
the Newton constant and the cosmological constant in the naive continuum
limit. The prefactor $1/C(T)$ is a remnant of the invariance group and 
divides out the internal symmetry factor of the triangulation. 
The free energy $F$ in the canonical ensemble 
is defined as:
\beq
e^{F(\k_2,N_4)} = 
\sum_{T \in {\cal T}(N_4)} \frac{1}{C(T)} e^{\k_2 N_2(T)}
\label{can}
\eeq
The sum runs over fixed topology $4d$ simplicial manifolds 
with a fixed number $N_4$ of $4$-simplices. 
In the large volume limit $N_4\rightarrow \infty$
the free energy is assumed to have  the form~:
\beq
F(\k_2,N_4) =  N_4 f(\k_2) + \delta(\k_2,N_4), 
\label{F}
\eeq
where the function $\delta$ is a finite size correction, 
{\em ie} for any $\k_2$ 
\beq
\lim_{N_4 \rightarrow \infty} \frac{\delta(\k_2,N_4)}{N_4} = 0 .
\label{Fn4}
\eeq

To recover some basic properties of the theory it is convenient
to study the derivatives of the free energy. 
The action density
\beq
r(\k_2,N_4) = \frac{1}{N_4} \frac{\pd F}{\pd \k_2} = 
\frac{\langle N_2 \rangle}{N_4}
\label{den}
\eeq
is normalized  such  that it becomes an intensive 
quantity in the thermodynamic limit $N_4 \rightarrow \infty$. 
It is related to the average bare
Regge curvature $R_{av}$ by $r = R_{av} + \alpha$ where 
$\alpha = \frac{10}{2\pi} \mbox{arccos}(1/4) \approx 2.09784$. 
The average is taken in the canonical ensemble (\ref{can}). 

The derivative with respect to $N_4$ 
\beq
\tilde{\k}_4(\k_2,N_4) = \frac{\pd F}{\pd N_4}  .
\label{k4}
\eeq
will be called the critical value of the parameter 
$\k_4$. Our definition differs slightly from the one 
proposed in  \cite{ckr}.
This quantity is a measure of finite size dependence of
the free energy, {\em ie}
how fast it approaches the thermodynamic limit
$\tilde{\k}_4 \rightarrow \tilde{\k}_4(\k _2)$ for $N_4 \rightarrow \infty $. 
In the thermodynamic limit $N_4 \rightarrow \infty$  the value
$\tilde{\k}_4 (\k _2)$ defines a critical line of the model corresponding 
to the radius of convergence of the series (\ref{can}).
 
Taking the second derivative, we see that  $\tilde{\k}_4$ is related to the 
action density $r$ in the following way~:
\beq
\frac{\pd \tilde{\k}_4}{\pd \k_2} = r + N_4 \frac{\pd r}{\pd N_4}
\label{dk4}
\eeq
It is important to note that in the large $N_4$ limit the second 
term on the right hand side of (\ref{dk4}) goes to zero, 
and the critical parameter $\tilde{\k}_4$ becomes 
an integral of $r$. Thus, if $r$ is independent of topology
in the thermodynamic limit, so is $\tilde{\k}_4$, unless
the integration constant depends on the topology.
To fix the integration constant, one just has to measure $\tilde{\k}_4$
for one particular value of $\k_2$.

\section{Methods}
The average action density (\ref{den}) can easily be measured in 
canonical simulations.
The quantity $\tilde{\k}_4$ (\ref{k4}) requires 
non-conventional methods. Some of them, like those based on 
the sum rules \cite{bk} or an analysis of the baby universe 
distributions \cite{ajt}, permit to directly extract the 
next-to-leading volume corrections. These methods are unfortunately
limited to the elongated phase. Following \cite{ckr}, we adopt here 
a  more general method based on 
multi canonical simulations which works equally well 
in the entire range of the coupling $\k_2$.
To learn about how the free energy depends on $N_4$ one
lets the volume  fluctuate in the external potential $U(N_4)$.
Measuring the resulting $N_4$ distribution and combining it
with the known form of $U$ one gets the dependence of the free energy 
on $N_4$.
The freedom one has in choosing of the potential $U$ should be
used to minimize the error for the quantity one wants to measure.
In this particular case one wants to measure $\pd F / \pd N_4$ 
for the function $F$, which is expected to smoothly approach 
a function  $N_4 f(\k _2)$ linear in $N_4$ for large $N_4$ as given 
in eq. (\ref{F}), (\ref{Fn4}). 
A Gaussian term controlling fluctuations around a fixed 
volume $V_4$ is well suited for this problem \cite{ckr}~:   
\beq
U = -\k_4 N_4 + \frac{\gamma}{2} (N_4 - V_4)^2
\label{u}
\eeq
but other terms can be used as well \cite{aj1}.
The multicanonical partition function for this potential
reads~: 
\beq
Z = \sum_{T}  e^{\k_2 N_2 + U(N_4)} =
\sum_{N_4} e^{F(\k_2,N_4) - \k_4 N_4 + \frac{\gamma}{2} (N_4 - V_4)^2}
\eeq
Expanding $F(\k_2,N_4)$ around $N_4 = V_4$ one  sees
that the $N_4$ distribution becomes  Gaussian~:
\beq
P(x = N_4 - V_4) \sim 
\exp( -\frac{\Gamma}{2} (x - x_0)^2 + ...)
\label{gauss}
\eeq
when $\k_4$ is tuned to be close to the derivative 
$\pd F / \pd V_4$ and  $\gamma$ is much larger than the
second derivative $\pd^2 F / \pd V_4^2$. This means that the
range of the distribution is much smaller than the typical scale
for changes in $F$. The parameters $\Gamma$ and $x_0$ are
related to $F$ and $U$~: 
\beq
\Gamma = \gamma - \pd^2 F / \pd V_4^2, \quad 
x_0 = \big( \pd F / \pd V_4 - \k_4 \big)/\Gamma.
\eeq
Both these quantities can be measured in the simulations, namely $\Gamma$ 
from the width of the distribution and $x_0$ from the
shift of the maximum from $N_4=V_4$. From the numerical results
one gets an estimator for the critical coupling  
at $V_4$~: 
\beq
\tilde{\k}_4 = \pd F/\pd V_4 = \Gamma x_0 + \k_4.
\eeq
In this formula, $\k _4$ is the coupling used in the simulation, from which
one extracted $x_0$ and $\Gamma $.  The estimator can
be improved by minimizing $x_0$. This can be done recursively by
setting $\k_4 = \tilde{\k}_4$ in the potential . 

After performing measurements one can justify the validity of
the assumptions used to write the Gaussian approximation 
(\ref{gauss}). The value for $\gamma $ should not be too large, because 
suppressing volume-fluctuations can spoil the mobility of the algorithm. 
On the other hand, it should not be too small either, because the average
number of sweeps between two configurations with canonical volume
$V_4$ grows with decreasing $\gamma $. In particular we checked that 
for $\gamma=0.0001$, $\Gamma$ computed from the width of the distribution, 
$\Gamma = 1/\sqrt{\sigma^2(N_4)}$, {\em i.e.} $\Delta N_4 = 100$, was equal
to $\gamma$ within the error bars, meaning that 
$\gamma \gg \pd^2 F / \pd^2 V_4$.
Therefore the free energy changes very slowly in the range 
of the distribution width as needed for the approximation (\ref{gauss}).

In this investigation we have performed simulations with
$V_4 =$ 4000, 8000, 16000, 32000, 64000.

Because the local moves \cite{gv92} used in the Monte Carlo simulation 
preserve the topology $t$ of the manifold, $t$ is given by the
topology of the starting configuration. In our simulation we choose to simulate
a spherical $S^4$ topology and two different tori, namely
$S^3\times S^1$ and  $S^1 \times S^1 \times S^1 \times S^1$.

The starting configuration for the sphere is, as usual, the $4d$-boundary
of a 5-simplex. The manifold $S^3 \times S^1$ can be produced from a
spherical manifold by taking away two separate 4-simplices
and gluing together the boundaries $b_{1}, b_{2}$, which are created when 
removing the two 4-simplices. However, two vertices in a 
4d simplicial manifold may have at most one common link. To avoid creation
of a double connection in the gluing procedure, the distance between the
vertices on $b_{1}$ to
those on $b_{2}$ has to be at least three links. One can ensure this without 
going through a tedious check, by gluing cyclically together three
copies of the original (spherical) manifold, {\em i.e.} 
$b_{1}/b_{2}', b_{1}' / b_{2}'', b_{1}''/b_{2}$. The (double-) prime is
used to distinguish the different copies.  The boundaries $b_1, b_2$ are 
chosen such that they have no common vertex.

The $S^1 \times S^1 \times S^1 \times S^1$ manifold can be built out
of the regular $3^4$ square torus by dividing 
each elementary $4d$-cube into 4-simplices 
in the following way. For each cube we mark two points 
$p_0=(0,0,0,0)$ and $p_4=(1,1,1,1)$ lying on the opposite ends of
the main diagonal and connect them by one of the shortest paths going
along the edges of the cube. The shortest path goes through
$4$ edges, each in a different direction, and through three points,
say $p_1,p_2,p_3$. There are $24$ such paths. Each set of points 
$p_0,p_1,p_2,p_3,p_4$ forms a 4-simplex.

\section{Finite Size Analysis and  Results}
The elongated phase of simplicial gravity is well
described as an ensemble of branched polymers \cite{aj2}, \cite{bi96},
\cite{bbpt96}.

This means that in  the large volume limit one can use the Ansatz
\beq 
F(\k_2,N_4) = N_4 f_0(\k_2) + (\gamma - 3) \log N_4  + f_1(\k_2) 
\label{frel}
\eeq
for the free energy. The corrections are expected to be of order
${\cal O}(1/N_{4})$. The correction coefficient
$\gamma$ is assumed to depend  only on the genus $g$ of the underlying 
branched polymer structure. 

Differentiating (\ref{frel}) with respect to  $\k_2$ one sees
that for the action density $r$ (\ref{den}) 
\beq
r = f_0'(\k_2 ) + \frac{ f_1'(\k_2 )}{N_4}
\eeq 
the logarithmic corrections disappear.
Therefore one should take the next corrections, namely $1/N_4$, 
into account. They can appear in $r$ for purely kinematic reasons.
To understand their origin consider the limit of large positive $\k_2$,
in which only triangulations maximizing $N_2$ contribute to 
the sum (\ref{can}). Such triangulations can be obtained 
from  barycentric subdivisions of 4-simplices applied
successively to a minimal starting configuration, possibly mixed with 
micro-canonical transformation, which do not change $N_2$ and $N_4$.
By the minimal configuration we mean the minimal volume triangulation 
which maximizes $N_2$. For the barycentric subdivisions one gets the relation
$N_2 = 5/2 N_4 + c^{0}$, where the constant 
\beq
c^{0} = N_{2}^0 - 5/2 N_{4}^0 
\eeq
characterizes the initial minimal configuration (indicated by the index 0).
The number $N_{2}$ of triangles is related to the
action density $r = N_2/N_4 = 5/2 + c^{0}/N_4$. 
This means, the constant $c^{0}$ leads to $1/N_4$ corrections of 
the action density. The contributions to the sum (\ref{can}) from 
triangulations built from non-minimal ones ({\em ie} smaller $N_2$) 
are suppressed exponentially by $\exp - \k _2 (c^0 - c)$, 
where $c$ characterizes the non-minimal start 
configuration with $c(\k _2) < c^0$.

For the sphere, the minimal configuration is the
surface of a $5$-simplex, and therefore $c^{0}$ is known. 
For the other topologies
we extracted the number listed below by a numerical experiment. 
With the standard topology preserving Monte Carlo process we use a cooling 
procedure, in which we increase $\kappa _4$, to decrease $N_4$, and $\k _2$
to maximize $N_2$. In fact we have to increase $\k _2$ slowly compared to
$\k _4$, because increasing $\kappa _2$ also increases the pseudo-critical
coupling $\tilde{\k}_4(\k _2)$.

For the topologies studied, we found that the 
following configurations are minimal~:
\beq
\begin{array}{rllll}
S^4 & : &
N_{4}^0 = 6, & N_{2}^0 = 20 & \Rightarrow c^{0}=5, \\ &&&& \\
S^3 \times S^1 & : & 
N_{4}^0 = 110, & N_{2}^0 = 44 & \Rightarrow c^{0}=0, \\ &&&& \\
S^1 \times S^1 \times S^1 \times S^1 & : &
N_{4}^0 = 704, & N_{2}^0 = 1472 & \Rightarrow c^{0}=288. \\
\label{minim}
\end{array}
\eeq
For the sphere $S^4$ the $1/N_4$ effect is very difficult to detect 
already for the volumes in the range of a few thousand 4-simplices 
and it would require extremely long runs to reduce the 
error bars below it.
For manifolds $S^3\times S^1$ the effect is not present at all. 
The corrections are, however, two orders of magnitude larger for 
$S^1 \times S^1 \times S^1 \times S^1$ and are
measurable in the volume range used in the simulations. 
This estimation of the $1/N_4$ effect is exact for infinite
positive $\k_2$ but one expects it to work although with a slowly
varying coefficient $c(\kappa _2)$ in the entire elongated phase. 

\begin{figure}[t]
\psfig{file=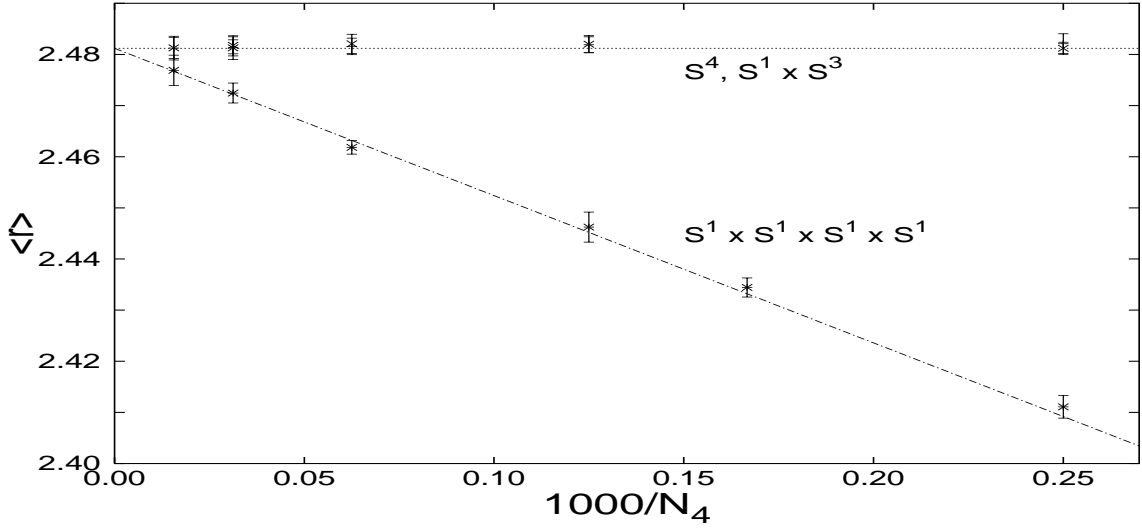,angle=270,width=11cm,height=10cm,rheight=7cm,rwidth=7cm}
\caption{The volume dependence of the action density $r$ for three different
topologies in the elongated phase at $\k_2 = 2.0$}
\label{vcr}
\end{figure}
In fig.\ref{vcr} we show the action density $r$ measured  
in the elongated phase for $\k_2=2.0$ against $1/N_4$. 
On the same figure we display the curve $r_\infty + c^0/N_4$ 
with $c^0=288$, and $r_\infty = 2.482$ which fits the data points very 
well. We note that that $r_\infty $ does not
depend on topology, at least for those used in the simulation. 

We find, with the statistics available, no volume- or topology-dependence
of $\tilde{\k_4}$. This is compatible with the Ansatz
\beq
\tilde{\k}_4(\k_2,N_4) =   f_0(\k_2 ) + \frac{ \gamma - 3}{N_4},
\label{k4elon}
\eeq
because $\gamma$ is known  to be ${\cal O}(1)$. 
With the method used, the correction could only be separated from the
statistical noise with an disproportionate amount of computer time. One could
instead use the methods used in \cite{ajt} to determine $\gamma$. 
We find  $\tilde{\k}_4^{\infty} = 5.659(4)$ for $\kappa _2 = 2.0$ in the
infinite volume limit. 

In the crumpled phase, we use the power-law Ansatz 
\beq 
F(\k_2, N_4) = N_4 f_0(\k_2) + f_1(\k_2)  N_4^{\delta }  
\eeq

\noindent
Taking the derivative with respect to $N_4$ one gets: 
\beq
\tilde{\k}_4(\k_2,N_4) =  f_0(\k_2 ) + 
     \delta f_1(\k_2) N_4 ^{ \delta - 1}.
\label{k4crump}
\eeq

\noindent
For the action density (\ref{den}) one finds
\beq
r = f_0'(\k_2 ) + f_1'(\k_2) N_4^{\delta - 1} 
\label{rcrump}
\eeq
\noindent
We checked the Ansatz by fitting our numerical data for
$r(N_{4})$ (\ref{rcrump}):
\begin{center}
\begin{tabular}{|r|c|c|c|c|}\hline
Topology &  $\delta $ & $r^{\infty} = f_{0}'$ & $\log(f_{1}')$&
$\chi^{2}/\mbox{dof}$\\\hline
$S^{4}$  & 0.5 (2) & 2.039 (+0.010 / -0.013)&1.9 (1.3) & 0.78\\
$(S^{1})^{4}$ & 0.6 (2) & 2.028 (+0.008 / -0.016) & 1.0 (0.9) & 0.87 \\
$S^{1}\times S^{3}$ & 0.5 (2) & 2.038 (+0.010 / -0.021) & 1.8 (1.3) & 0.31 \\
\hline
\end{tabular}
\end{center}
and for $ \k _{4}(N_{4})$ to (\ref{k4crump}):
\begin{center}
\begin{tabular}{|r|c|c|c|c|}\hline
Topology &  $\delta $ & $\k_4^{\infty}=f_{0}$&$\log(\delta f_{1})$&$\chi^{2}/\mbox{dof}$\\
\hline
$S^{4}$  & 0.6 (2) & 1.20 (2)& 1.4 (1.3) & 0.22\\
$(S^{1})^{4}$ & 0.6 (3) & 1.20 (2) & 1.4 (1.5) & 0.10 \\
$S^{1}\times S^{3}$ & 0.6 (2) & 1.20 (2) & 1.4 (1.3) & 0.18 \\
\hline
\end{tabular}
\end{center}

\begin{figure}
\psfig{file=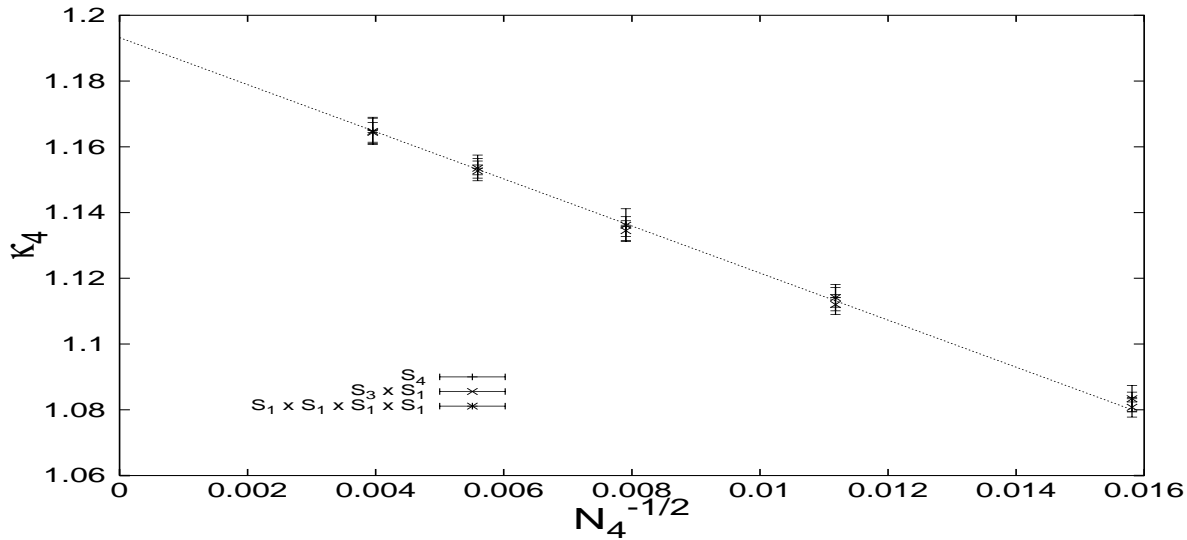,angle=270,width=11cm,height=10cm,rheight=7cm,rwidth=7cm}
\caption{The volume dependence of the critical $\k _4$ for 
three different topologies in the crumpled phase at $\k_2 = 0.0$}
\label{k4cr}
\end{figure}

In the tables we give the logarithms of $f_1'$ and $\delta f_1$, while 
these have approximately symmetric errors.
One can see that the the values do not depend, within errors, on  
topology. In figure \ref{k4cr} we show the numerical data for $\k _{4}$ for
the three topologies and $\delta = 0.5$. 

The value $\delta = 0.5$ can 
be understood by looking at the typical configurations, that dominate
the ensemble for large negative $\kappa _2$. These are configurations,
which minimize the free energy, {\em i.e.} which have for a fixed 
volume the minimal number of vertices and thereby the minimal number
$N_2$ of triangles. For the three-dimensional case
such configurations were constructed in \cite{d}. This construction
can easily be extended to the four-dimensional case.
One starts with a $2d$ triangulation of the sphere with $t$ triangles. 
At each triangle one builds a $4d$-pancake neighborhood from $q$
4-simplices lying around the triangle in such a way that the 
links opposite to this triangle form a circle. An opposite link
is defined to be a link, which does not share a common vertex
with the triangle. The next step is to put neighboring 
pancakes together by identifying these circles with
the $3d$ faces of the neighboring pancakes.
Each pancake has three such faces which lie between the
circle and an edge of the basic triangle. It also has
three neighboring pancakes. After this step one gets an $S^4$ 
sphere with $t*q$ 4-simplices and $2+t/2+q$ vertices. 
The highest connectivity $N_4 \sim (N_0-2)^2 / 2$ is reached 
when $t=2q$. The number $N_2$ of triangles is a linear combination of
$N_0 $ and $N_4$. Using equation \ref{den}  one finds 
$r \propto 1/ /sqrt (N_4)$ {\em ie.}  $\delta = 0.5$ 
for these configurations.
We want to note, that in \cite{aj3} an argument in favor of
$\delta = 0.75$ was given. This can not be excluded by our numerical
data.

In \cite{bbkp96} evidence was given, that for spherical topology the 
phase transition is of first order. This was confirmed in
\cite{b96}, \cite{ctkr96}. We could observe flip-flops in the
action density also for the two tori under investigation.
We interpret this as a hint, that the transition is of first order in these
cases as well.

Finally we want to comment on the behavior of the algorithm in the
crumpled phase. The typical configuration  of simplicial gravity in this
regime has one so-called singular link. The local volume of a singular link, 
{\em i.e.} the number
of four-simplices which contain this link, diverges, when the volume
of the entire configuration goes to infinity. The relaxation time, 
{\em i.e.} the number of Monte Carlo sweeps required until the
singular link appeared in the configuration, was extremely long for 
all topologies. This effect is even more pronounced if one
starts with a branched-polymer like configuration. For the 
$(S^1)^4$-torus  and small volumes (less than 64k 4-Simplices) the situation 
is even worse.  It was, at least for the run-length used in our 
numerical experiments, impossible to relax to such singular configurations. 
On the other hand, we know that they exist, because they could be
reached by shrinking down larger configurations containing a singular link.
It seems it is difficult for the algorithm to deal with two different
defects, the singular link and the hole, at the same time.
We will discuss this point, which is directly related to the question of 
practical ergodicity, more carefully in a forthcoming publication. 

\section{Discussion and Conclusions}

In this paper we have investigated the behavior of the
entropy density and the curvature for three different
topologies in four dimensional simplicial gravity.
We employed manifolds consisting of between 4000
and 64000 simplices. We concentrated on two values of
the gravitational coupling constant $\kappa_2$,
one in the crumpled and one in the branched polymer 
phase. We found that in both the cases the value of the
entropy density and curvature in the infinite volume
limit are equal for these three topologies. This gives
further support to the conjecture that these limits exist,
a question which was discussed in the literature some time ago
\cite{ckr94}, \cite{aj3}, \cite{bm95}. 

Furthermore, it gives  support to the hypothesis that
all topologies contribute to a sum over topologies,
like in two dimensions. The value of the entropy exponent
can be compared with the  estimates given
in \cite{bbcm94}, based on a summation over all distributions
of curvature. This estimate, which surprisingly enough is exact 
for the leading term in two dimensions, does, however, not directly
give results in agreement with our numerical data for $\tilde{\k} _4$.

We further analyzed in detail the finite size effects,
and compared them to estimates from simple kinematic
arguments. This approach explain very well the finite
size corrections. It may, of course still be possible
that those effects are of a more complicated nature
near the transition.

Finally, as also seen in previous investigations,
we observe that the algorithm has very long relaxation
times in the crumpled phase. We even found that for the
$(S_1)^4$ torus, the ground state seemed not to accessible
using the approximately fixed $N_4$ algorithm, but only by
passing through states with much larger $N_4$-values.
This breakdown of practical ergodicity is still under
investigation. 

\section{Acknowledgments}
We are grateful to J. Ambj\o rn, P. Bia\l{}as, 
and J. Jurkiewicz and A. Krzywicki 
for discussions. The work was supported by the Deutsche 
Forschungsgemeinschaft under grant Pe340/3. We thank the HLRZ J\"ulich,
for the computer time on the Paragon on which the work was done.

\end{document}